\documentclass{ifacconf}

\usepackage{graphicx}      
\usepackage{natbib}        
\usepackage{graphics} 
\usepackage{amsmath}
\usepackage{amssymb}  
\usepackage{amsfonts} 
\usepackage{graphicx,float,wrapfig,subfigure}
\usepackage{graphicx}
\usepackage{multirow}
\usepackage{adjustbox}
\usepackage{soul,color}
\usepackage{url}
\usepackage{comment}
\usepackage{enumitem}

\begin{document}
\begin{frontmatter}

\title{Data-driven Energy Management Strategy for Plug-in Hybrid Electric Vehicles with Real-World Trip Information}


\author[First]{Yongkeun Choi} 
\author[First]{Jacopo Guanetti} 
\author[Second]{Scott Moura}
\author[First]{Francesco Borrelli}

\address[First]{Model Predictive Control Lab, Department of Mechanical Engineering, 
   University of California, Berkeley, CA 94720 USA (e-mail: yk90@berkeley.edu).}
\address[Second]{Energy, Controls, and Applications Lab, Department of Civil and Environmental Engineering, 
   University of California, Berkeley, CA 94720 USA}

\begin{abstract}                
This paper presents a data-driven supervisory energy management strategy (EMS) for plug-in hybrid electric vehicles which leverages Vehicle-to-Cloud connectivity to increase energy efficiency by learning control policies from completed trips.
The proposed EMS consists of two layers, a \textit{cloud} layer and an \textit{on-board} layer. The cloud layer has two main tasks: the first task is to learn EMS policy parameters from historical trip data, and the second task is to provide the policy parameters along a certain route requested from the vehicle. 
The on-board layer receives the learned policy parameters from the cloud layer and computes a real-time solution to the powertrain energy management problem, using a model predictive control scheme.
The proposed EMS is evaluated on more than 3000 miles (48 independent driving cycles) of real-world trip data, collected along three commuting routes in California. For the routes, the proposed algorithm shows 3.3\%, 7.3\%, and 6.5\% improvement in average MPGe when compared to a baseline EMS.
\end{abstract}


\end{frontmatter}

\section{Introduction}
\label{section : introduction}

Plug-in hybrid electric vehicles (PHEVs) implement an energy management system (EMS), that in real-time allocates the current power demand (from the driver or the longitudinal control, and from the auxiliary systems) to the on-board power sources.
A primary goal in EMS design is energy efficiency. 
This is achieved by balancing the use of fuel and electric energy in order to maximize trip-wise efficiency.
A major issue is that the electric energy stored on-board is limited and battery recharge is time-consuming. 
As a consequence, an optimal EMS policy requires perfect knowledge of the future power demand and charging opportunities, throughout the trip.
In practice, accurate forecasts can be expensive or hard to get, and require user involvement -- for instance in planning the route and the stops at charging stations.

Therefore, most commercial PHEVs implement the so-called Charge Depleting-Charge Sustaining (CD-CS) energy management strategy. 
In CD-CS, vehicles mostly utilize electric-only mode (CD phase) until the battery state of charge (SOC) reaches a minimum limit.
Afterwards, they consume fuel more aggressively to maintain SOC near the minimum, and provide propulsion (CS phase).
An intuitive motivation for the CD-CS strategy is the fact that running the vehicle on electricity is generally more energy-efficient than on fuel.

Most systematic approaches to design an EMS are based on optimal and predictive control ideas. 
While there is no analytic solution to this problem, many numerical approaches have been proposed. 
These methods show that the CD-CS strategy explained above is, in general, sub-optimal. 
The optimal policy generally uses both fuel and electric power in a blended manner throughout the trip, using the knowledge of the power demand, see e.g. \cite{Sharer2008Plug, Stockar2009Energy}.
Since in practice the power demand is only known in real-time, a new approach is required to deliver optimal strategies in production vehicles.

A popular approach which tries to address this problem is to use a locally optimal control policy, and to expose one or more tuning knobs (usually weights in the cost function). 
Various adaptation and estimation approaches are used to modulate the knobs in real-time, see e.g. \cite{Stockar2011Energy, Musardo2005AECMS, Serrao2011Comparative, Manzie2015State, Guanetti2016}.
While these methods can give near-optimal results in some driving conditions, simulations or experiments in a variety of driving conditions are required to empirically tune the controller and verify it achieves satisfactory performance.

Another possible approach is to build a stochastic model of the power demand, for instance by training a Markov chain using historical profiles of the power demand.
Finding an optimal EMS policy then becomes a stochastic optimal control problem, which can be solved by stochastic dynamic programming (SDP) techniques as in \cite{Moura2011Stochastic, Opila2012Energy, Johannesson2007Assessing}.
While the computational effort required to solve SDP is large, this computation is only performed once and offline. 
In real-time, the optimal policy can be stored in the form of a lookup table.
On the other hand, this workflow is inconvenient when the model needs to be retrained frequently, for instance because new driving data are available.
A way around this issue is to incorporate a learning mechanism, in such a way that a personalized policy can be learned over time.
This can be achieved via stochastic model predictive control (SMPC).
Because the SMPC problem is solved in real-time, the stochastic part of the model can also be learned in real-time, as driving data become available, see \cite{Cairano2014Stochastic}.
A limitation of this approach is the large computational effort required in real-time.

To mitigate such large computational effort, approximate dynamic programming (ADP) can be incorporated in powertrain control.
The recent work \cite{Qi2016Data} investigates this possibility for a PHEV. 
More specifically, the algorithm proposed therein attempts to learn the optimal EMS policy using data of vehicle speed, road grade, battery charge, power demand, and availability of charging stations.
ADP is appealing because most of the computational effort is moved offline, and learning can be naturally incorporated.
However, in \cite{Qi2016Data}, synthesized trip data sets are used. More specifically, the power demand is estimated using a longitudinal vehicle model. 
However, the uncertainty associated with the longitudinal model may degrade performance.
In this paper, we present a data-driven supervisory EMS for PHEVs that aims at improving the real-world PHEV energy efficiency via Vehicle-to-Cloud (V2C) connectivity, by utilizing real-world trip data sets.

The main contribution of this paper is twofold. 
First, we propose a two-layer EMS framework that systematically provides a real-time, near-optimal solution using historical data via V2C connectivity. 
Second, we utilize more than 3000 miles of real-world driving data, that include not only the GPS and speed traces, but also traffic conditions and accurate measurements of the vehicle energy consumption, including auxiliary power, fuel flow, battery voltage and current, and speed and torque readings from the on board machines. 
This increases the fidelity of our simulation results.
We show that the proposed algorithm achieves, on three selected routes, 3.3\%, 7.3\%, and 6.5\% improvement in average MPGe when compared to a baseline EMS, and 1.3\%, 2.6\%, and 2.0\% deterioration of average MPGe when compared to the optimal, non-causal EMS computed by dynamic programming (DP).

This paper is structured as follows.
Section \ref{section : Modeling} describes a PHEV powertrain model used in this study.
Section \ref{section : controller design} presents the overall architecture of the proposed controller and explains each control layer in detail.
Section \ref{section : realworld driving data and model validation} explains the collected real-world driving data and how model validation is performed.
In Section \ref{section : result}, we provide simulation results based on the proposed EMS.
Section \ref{section : conclusion} discusses the implication of this research and addresses some directions for the future works.

\section{Powertrain Model}\label{section : Modeling}
\begin{figure}
    \centering
    \includegraphics[width=1\columnwidth]{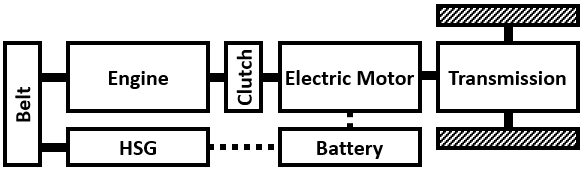}
    \caption{PHEV powertrain architecture}
    \label{fig : PTmodel}
\end{figure}

The powertrain architecture used in this study is a pre-transmission parallel hybrid as shown in Fig. \ref{fig : PTmodel}. 
It consists of an internal combustion engine, electric motor, hybrid starter generator (HSG), mechanical belt, clutch unit, transmission unit, and high voltage battery. 
Solid connections between components in Fig. \ref{fig : PTmodel} represent mechanical connections and dotted connections represent electrical connections.
In this section, the powertrain model used for control design and simulations is introduced. 

In the powertrain model, we make the following assumptions: (i) the transient actions of engaging the clutch and starting up the engine are simultaneous and instantaneous; (ii) torque responses of the engine and motor and gear shifts are instantaneous; (iii) the HSG is only used to start the engine (i.e. the HSG draws electrical energy whenever the engine state is switching from \textit{off} to \textit{on}).

At time $t$, Euler-discretized state equations with sampling time $t_s$ can be defined as
\begin{subequations}\label{eq : state equations}
\begin{align}
    \mathrm{SOC}(k+1|t) &= \mathrm{SOC}(k|t) 
    \nonumber \\ - t_{s} & \frac{V_\mathrm{oc}(k|t)-\sqrt{V_\mathrm{oc}^2(k|t) - 4R_b(k|t) P_b(k|t)}}{2R_b(k|t) Q_b}, \label{eq : SOC state dynamics}\\
    e_\mathrm{state}(k+1|t) & = e_\mathrm{switch}(k|t), \label{eq : engine state dynamics}
\end{align}
\end{subequations}
where $\mathrm{SOC}$ is the battery state of charge; $V_\mathrm{oc}$ and $R_b$ are the battery open-circuit voltage and internal resistance, respectively; both are nonlinear functions of the SOC; $P_b$ is the battery power; $Q_b$ is the battery capacity; both $e_\mathrm{state}$ and $e_\mathrm{switch}$ are boolean variables, which are the engine status and switch, true when the engine is on and commanded to be on, respectively.
More details on the model above can be found in \cite{Guzzella2013Vehicle}. 

The electrical battery power can be expressed as follows:  
\begin{equation}\label{eq : battery power equation}
    P_{b}(k|t) = P_{m}(k|t) + P_\mathrm{HSG}(k|t) + P_\mathrm{aux}(k|t),
\end{equation}
where $P_{m}$, $P_\mathrm{HSG}$, and $P_\mathrm{aux}$ denote electric motor power, HSG power, and auxiliary electric loads, respectively. 
The electric motor power $P_{m}$ is computed as
\begin{equation}\label{eq : electric motor power equation}
    P_{m}(k|t) = \frac{T_m(k|t) \omega_m(k|t)}{\eta_m(T_m(k|t),\omega_m(k|t))},
\end{equation}
where $T_m$ is the electric motor torque, $\omega_m$ is the electric motor speed, $\eta_m$ is the electric motor efficiency (a nonlinear function of $T_m$ and $\omega_m$).
While $P_\mathrm{HSG}$ can take both positive and negative values, in this work we restrict it to the positive domain; in particular, the machine only operates as a starter for the internal combustion engine, drawing a constant power from the battery for a half of a second at every engine startup event.

The \textit{internal} battery power, $P_q$, is used in this study, and it is the time derivative of the battery energy as defined in the work \cite{Murgovski2012Convex}. 
The engine power $P_{f}$ and the corresponding fuel rate $\dot{m}_{f}$ are given by
\begin{subequations}
\begin{align}
    P_{f}(k|t) & = \frac{T_e(k|t) \omega_e(k|t)}{\eta_e(T_e(k|t),\omega_e(k|t))} e_\mathrm{state}(k|t), \label{eq : fuel power equation}\\
    \dot{m}_{f}(k|t) & = \frac{P_{f}(k|t)}{\lambda_{f}},
\end{align}
\end{subequations}
where $T_e$ and $\omega_e$ are the engine torque and engine speed, $\eta_e$ is the engine efficiency (a nonlinear function of $T_e$ and $\omega_e$), $\lambda_f$ is the fuel lower heating value. 
From assumption (i), $\omega_m = \omega_e$ when the clutch is engaged (i.e., $e_\mathrm{state}(k|t) = 1$).

The traction torque is modeled as
\begin{equation}\label{eq : traction torque equation}
T_d(k|t) = g_{i}(k|t)\eta_{t}(T_m(k|t) + c_\mathrm{on}\eta_{c}T_e(k|t)),
\end{equation}
where $T_d$ denotes the demanded torque at the wheel. 
Symbols $\eta_{t}$ and $\eta_c$ denote the transmission and clutch efficiency, respectively; $c_\mathrm{on}$ is a boolean variable, true when the clutch is closed; $g_i$, $i = 1, ..., 6$ is the gear ratio corresponding to the $i^{th}$ gear number. 
In the powertrain model, $\eta_{t}$ and $\eta_c$ are assumed to be constant, and $c_\mathrm{on}$ is the same variable as $e_\mathrm{state}(k|t)$ based on assumption (i).

In summary, the states and inputs for the powertrain model are $x = [\mathrm{SOC} \quad e_\mathrm{state}]^{T}$ and $u = [T_{e} \quad e_\mathrm{switch}]^{T}$, respectively, and the disturbances are $w = [P_\mathrm{aux} \quad T_d]^{T}$.

\section{Controller Design}\label{section : controller design}

\begin{figure}[!t]
    \centering
    \includegraphics[width=0.5\linewidth]{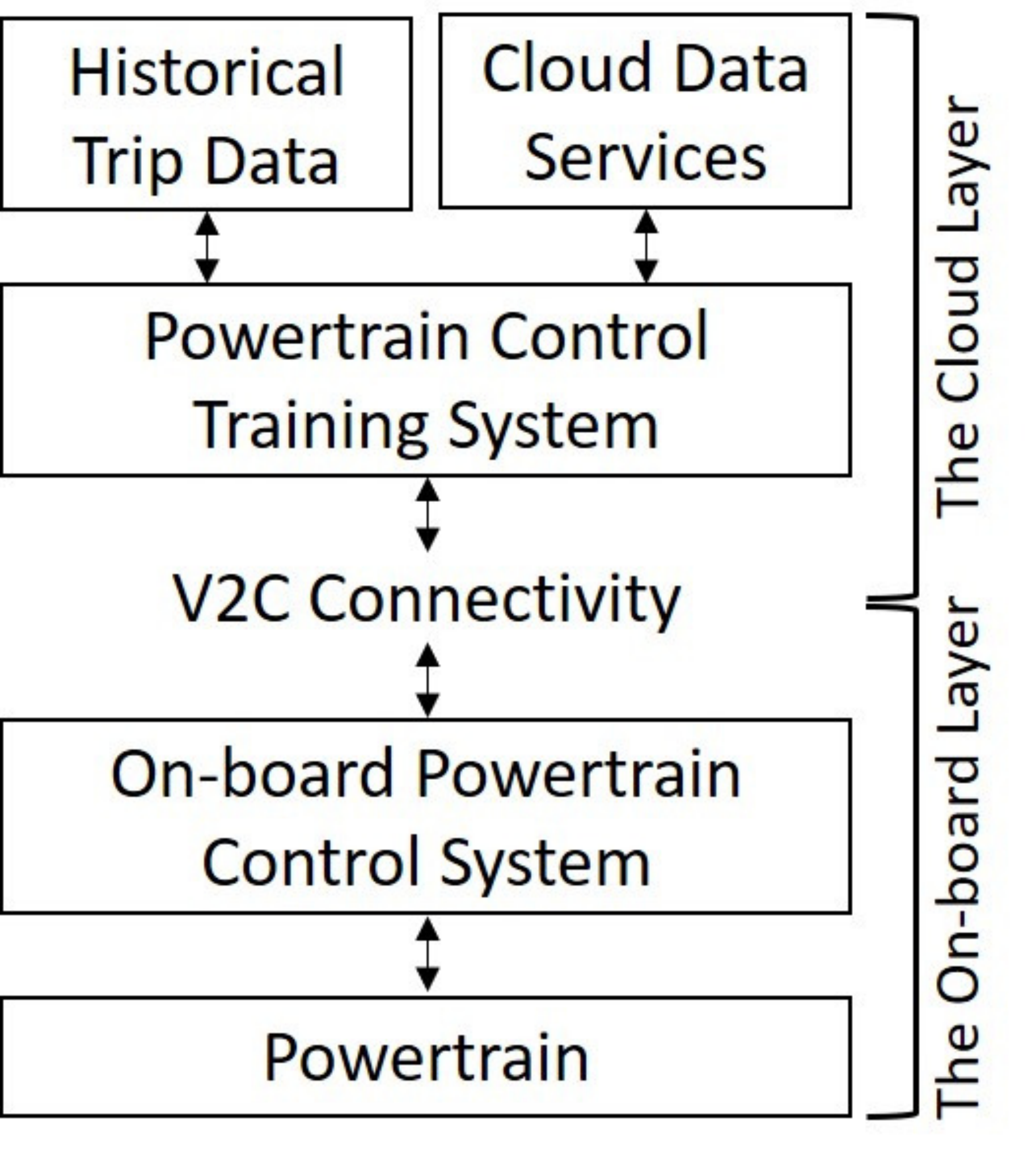}
    \caption{Overall architecture of the proposed controller.}
    \label{fig : OverallArchitcture}
\end{figure}

The overall architecture of the proposed EMS design is shown in Fig. \ref{fig : OverallArchitcture}.
In the cloud layer, the powertrain control training system computes the optimal value function for each historical trip by DP.
From cloud data services, real-time traffic forecasts are sent to the powertrain control training system.
Using the historical optimal value functions and the real-time forecasts, control parameters, which approximate the computed optimal value functions, are computed with a supervised learning algorithm.
Finally, the resulting control parameters are sent to the on-board layer via V2C connectivity.
In the on-board layer, the real-time powertrain control problem is solved using an MPC scheme using an approximated value function as a compressed representation of long-term information about the rest of the trip. 
This way, the on-board real-time powertrain control system can be long-sighted while only performing low-complexity local optimization in real-time.
The control parameters can be updated along the route continuously.

\subsection{Offline computation with Dynamic Programming (The Cloud Layer)}

The objective of the EMS for PHEVs is energy efficiency. 
This is achieved by minimizing the total energy consumption for the overall trip. 
Miles Per Gallon equivalent (MPGe) is used as the metric to evaluate energy consumption in this work \cite{EPA2016New}.
Every time the cloud layer receives the data for a completed trip, it solves the following discrete deterministic optimal control problem using the Bellman realization (\cite{Bertsekas2017DP1}).

\begin{subequations}
\begin{alignat}{2}
    & \underset{u(0|0),...,u(N_{f}-1|0)}{\text{min}} \sum_{k=0}^{N_{f} - 1} c(x(k|0),u(k|0),w(k|0)), \label{eq : cost functions}\\
        & \text{s.t.} \quad \quad x(0|0) = x_0, \\
        & \quad \quad \quad \ x(k+1|0)= f(x(k|0),u(k|0),w(k|0)), \\
        & \quad \quad \quad \  0 = h(x(k|0),u(k|0),w(k|0)), \\
        & \quad \quad \quad \  x(k|0) \in \textbf{X}, u(k|0) \in \textbf{U}, \\
        & \quad \quad \quad \  \forall k \in [0,1,...,N_{f}-1],
    \end{alignat}
\end{subequations}

where the objective function $c$ is the sum of fuel power $P_f$ and battery internal power $Pq$ converted into gallon equivalent unit, $x_0$ represents the initial state of the powertrain, i.e., SOC and an engine state, $N_{f}$ is the final time step, $f$ represents the battery SOC and the engine dynamics as shown in (\ref{eq : SOC state dynamics} - \ref{eq : electric motor power equation}), $h$ represents traction torque constraints as shown in \eqref{eq : traction torque equation}.
The state constraint $\textbf{X}$ is expressed as
\begin{equation}
    \mathrm{SOC}^{\text{min}} \leq \mathrm{SOC}(k|0) \leq \mathrm{SOC}^{\text{max}}, 
\end{equation}
where $\mathrm{SOC}^{\text{min}}$ and $\mathrm{SOC}^{\text{max}}$ are predefined parameters. The input constraint, $\textbf{U}$, can be expressed as
\begin{equation}
    T_{e}^{\text{min}}(k|0) \leq T_{e}(k|0) \leq T_{e}^{\text{max}}(k|0),    
\end{equation}
where the torque bounds $T_{e}^{\text{min}}$ and $T_{e}^{\text{max}}$ depend on the axle speed.
In (\ref{eq : cost functions}), $P_f$ represents the fuel power as defined in \ref{eq : fuel power equation} and $P_q$ represents the \textit{internal} battery power as described in section \ref{section : Modeling}.

Then the following Bellman equation is solved recursively backwards in discrete time.
\begin{align}
    V_k^*(x(k|0)) = \underset{u(k|0) \in \textbf{U}}{\text{min}} & \ c(x(k|0),u(k|0),w(k|0)) \nonumber \\ & + V^*_{k+1}(f(x(k|0),u(k|0),w(k|0))),
\end{align}
where $V^*_{k+1}$ is the optimal value function at the next time step. 
The resulting optimal value functions along with granular trip data are saved in the cloud layer.

\subsection{Approximating Value Function (The Cloud Layer)}\label{subsection : approx val func}
The computed value functions and the trip data are used to train control parameters. 
To approximate the value functions, a supervised learning algorithm, specifically a linear regression is used as follows:
\begin{equation}\label{eq : fitting_DP}
    r(k|0) = \underset{r}{\text{argmin}} \ (\Hat{V}_k(x^{s}(k|0),r) - V_k^{*}(x(k|0)))^{2}, 
\end{equation}
where
\begin{equation}\label{eq : fitting_DP_sub}
    \hat{V}_k(x^{s}(k|0),r(k|0)) = \sum_{l=1}^{m} r_{l}(k|0) \cdot x_{l}^{s}(k|0).
\end{equation}
$x^{s}$ represents feature states (which will be described later in this section); $\hat{V}_k$ is an approximated value function; $r$ is a parameter (weight) vector; $x_{l}^{s}$ and $r_{l}$ are the $l^{th}$ component of $x^{s}$ and $r$, respectively.
$k$ is position to handle different travel duration of different trips on the same route.

The feature states $x^{s}$ include powertrain states (SOC and engine status), vehicle information (such as vehicle speed), and environment features (such as an estimated time left until the destination). 
The quality of the approximation $\hat{V}_k(x^{s},r)$ depends on the choice of feature states. 
The following 7 feature states have been considered in this study, because they can be considered as driving pattern factors that affect energy usage \cite{Ericsson2001Independent}: SOC, engine status, average auxiliary electric loads, fuel consumption, average speed, average acceleration, and estimated time left.


\subsection{On-board Powertrain Control (The On-board Layer)}
In the on-board layer, a MPC scheme is used to solve the real-time powertrain control problem. 
At each time step, $t$, at the on-board layer, the following MPC problem of prediction horizon $N_h$ is solved and the first element of the resulting input sequence, $u(0|t)$, is applied to the system.
\begin{subequations}
    \begin{align}
        & \underset{u(0|t),...,u(N_{h}-1|t)}{\text{min}} \sum_{k=0}^{N_{h} - 1} \bigg\{c(x(k|t),u(k|t),w(k|t)) \nonumber \\ & \quad \quad \quad \quad \quad \quad \quad \quad \quad + \Hat{V}_{k+1}(x^{s}(k+1|t),r(k+1|t))\bigg\}, \label{eq : 1stepMPC_term_constMPC} \\
        & \text{s.t.}  \quad \quad x(k|t) = x_t, w(k|t) = w_t, r(k+1|t) = r_t, \label{eq : 1stepMPC_initial_constMPC} \\ 
        & \quad \quad \quad \ x(k+1|t) = f(x(k|t),u(k|t),w(k|t)), \label{eq : 1stepMPC_dynamics_constMPC} \\
        & \quad \quad \quad \ 0 = h(x(k|t),u(k|t),w(k|t)), \label{eq : 1stepMPC_equality_constMPC} \\
        & \quad \quad \quad \ x(k+1|t) \in \textbf{X}, u(k|t) \in \textbf{U}, \label{eq : 1stepMPC_stateinput_setMPC} \\
        & \quad \quad \quad \ k \in [0, 1, ..., N_h-1], \label{eq : 1stepMPC_stateinput_setMPC}
    \end{align}
\end{subequations}
where $x_t$, $w_t$, and $r_{t}$ are given states, disturbances, and a control parameter vector at each time step, t.

\section{Real-World Driving Data and Model Validation}
\label{section : realworld driving data and model validation}
In this section we briefly describe the real-world trip data used in the next section to validate the proposed data-driven EMS.
The trip data sets were collected from two identical test vehicles on three different routes in California.
The three (round-trip) routes are: (i) Fremont to Berkeley, (ii) Irvine to Chino, and (iii) Highland to Chino, and hereafter are labeled as \textit{Fremont route} (65 miles), \textit{Irvine route} (75 miles), and \textit{Highland route} (71 miles), respectively.
Each route is associated with 16 trips, yielding about 3400 miles of data.
The trip data includes GPS traces (GNSS 2.5 meters accuracy, according to \cite{GPSMK5OBU}) and measurements from vehicle sensors and estimators.
In particular, the measurements include the speed and torque of the gear box input shaft, the electric motor, the internal combustion engine torque, and the hybrid starter generator.
While the torque signals are generally not directly measured, but only estimated from other measured signals, the availability of granular speed and torque data allows to more directly evaluate how the proposed EMS impacts powertrain efficiency.

The sensors include a high accuracy fuel flow meter and an array of high accuracy current sensors (measuring the current of the air conditioning unit, the high voltage battery, the low voltage battery, and the on-board data acquisition system), that were installed on the test vehicles to precisely measure the energy consumption.
All the real-world trip data sets are collected from a powertrain system that implements a CD-CS strategy. 
Its measured energy consumption is used as the baseline in this study.

\begin{figure}[]
    \centering
    \subfigure[]{\includegraphics[width=1\linewidth, height=0.5\linewidth]{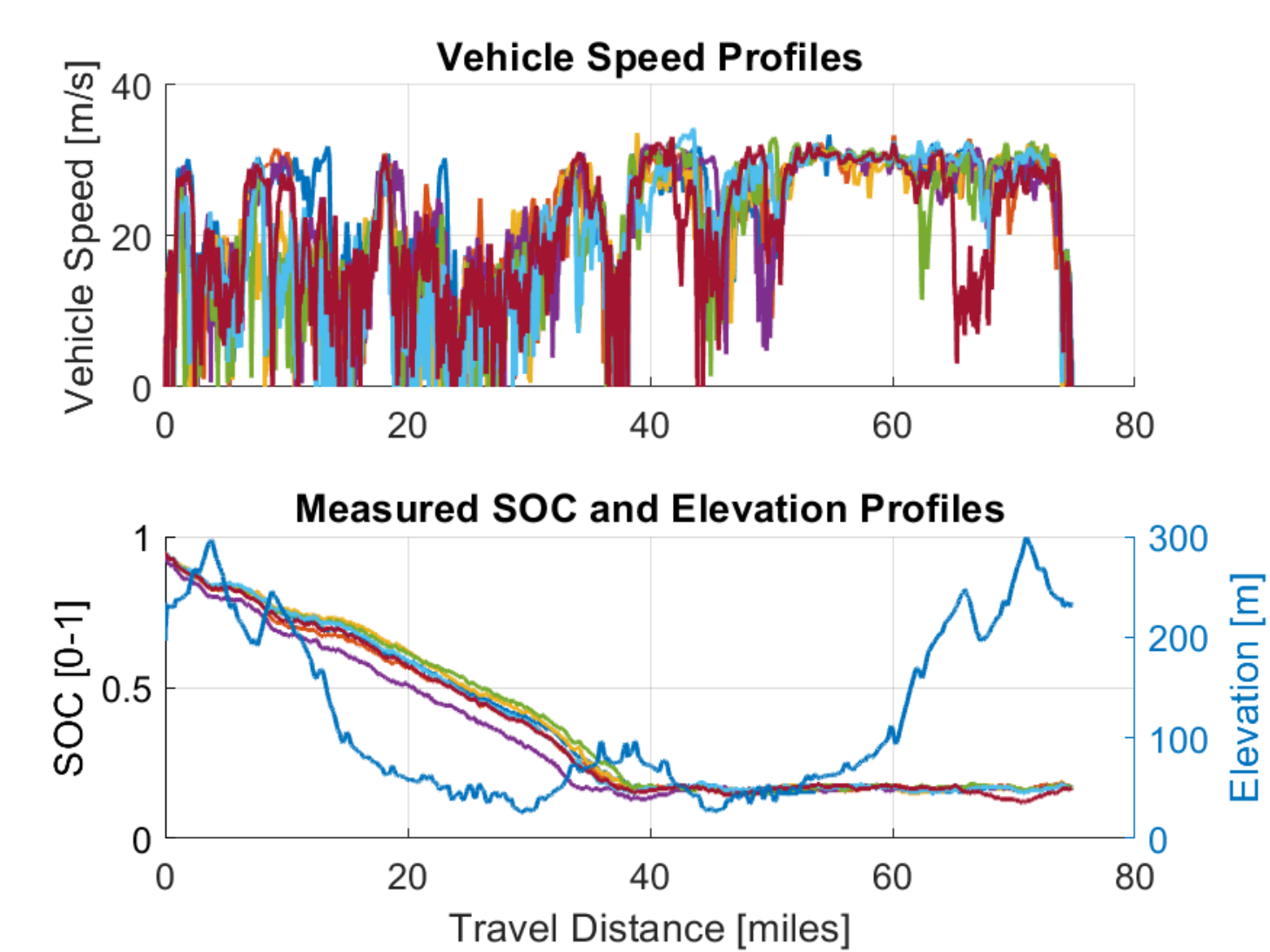}} 
    \subfigure[]{\includegraphics[width=1\linewidth, height=0.5\linewidth]{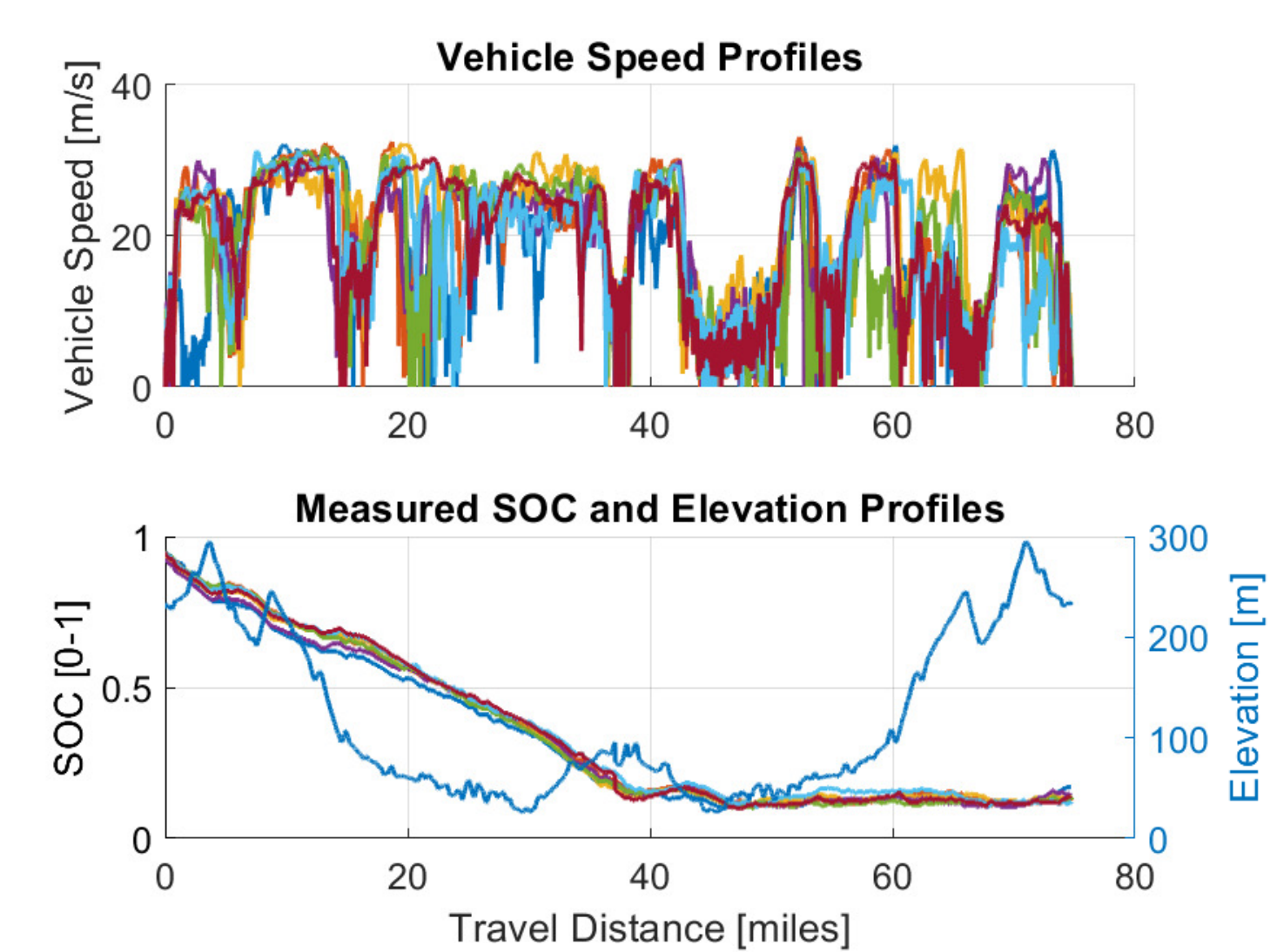}} 
   \caption{Measurements of vehicle speed (top), battery SOC and terrain elevation (bottom) collected on the Irvine route, showing the effect of (a) morning traffic and (b) afternoon traffic.}
    \label{fig : Irivne_two different driving}
\end{figure}

Figure~\ref{fig : Irivne_two different driving} shows samples of measured speed profiles, battery SOC discharge profiles (CD-CS), and terrain elevation profiles on the Irvine route.
Two different vehicle speed patterns are observed, due to the different traffic conditions at different times of the day. 
Morning data sets are collected between 7 am and 10 am, and afternoon data sets are collected between 3 pm and 6 pm.




\begin{figure*}
    \centering
    \subfigure[]{\includegraphics[width=0.32\linewidth ,height=0.25\linewidth]{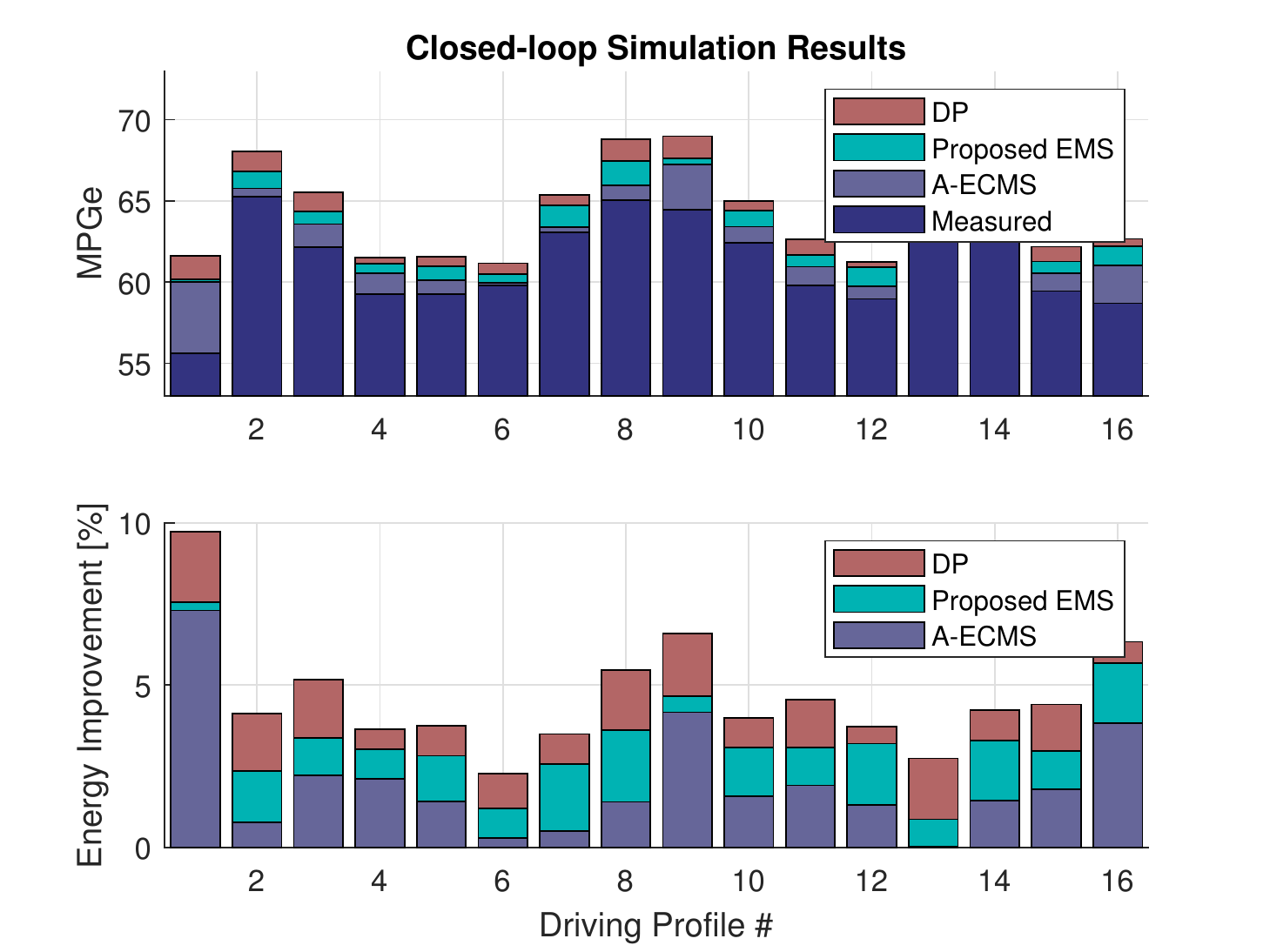}} 
    \subfigure[]{\includegraphics[width=0.32\linewidth, height=0.25\linewidth]{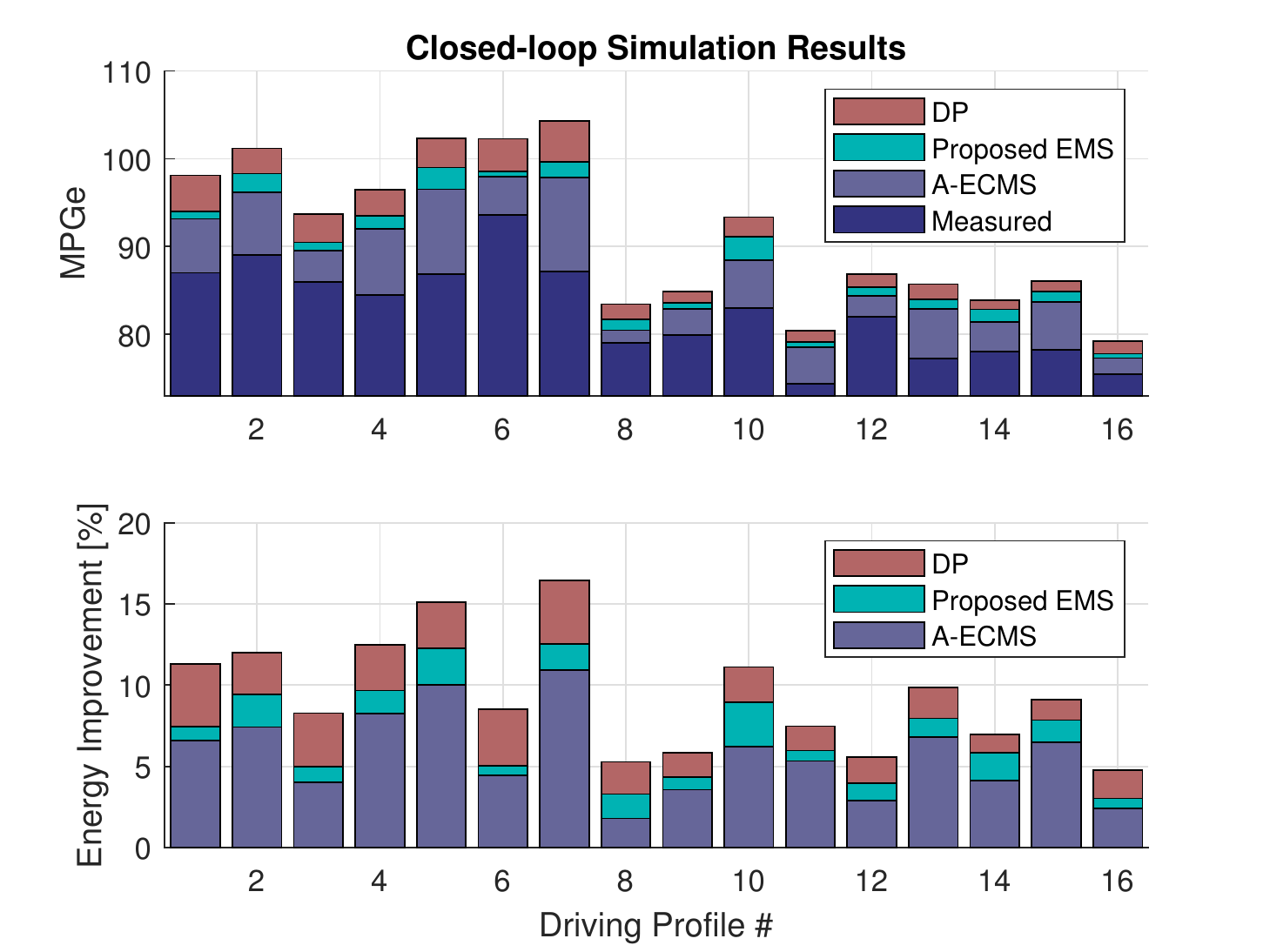}} 
    \subfigure[]{\includegraphics[width=0.32\linewidth, height=0.25\linewidth]{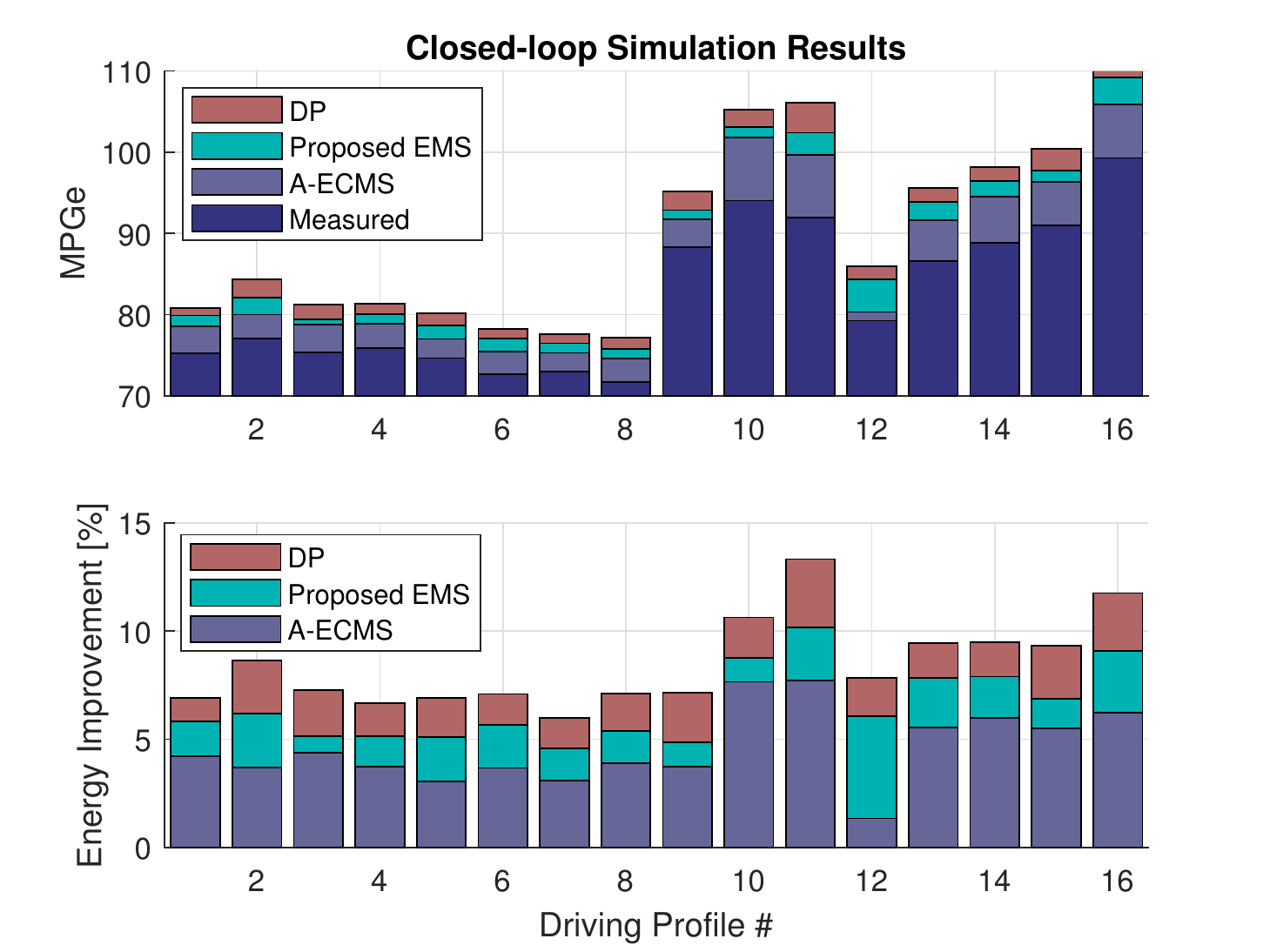}}
    \caption{The proposed EMS simulation results for (a) Fremont, (b) Irvine, and (c) Highland routes.}
    \label{fig : Sim_Results}
\end{figure*}

In this work, MATLAB is used to simulate the proposed controller and the powertrain dynamics with the mathematical models as described in Section \ref{section : Modeling}. Most of the constants and nonlinear functions used in the powertrain mathematical model are provided from the manufacturer of the vehicle, and the ones not given or not matched are fitted using a regression utilizing our granular data sets. 
This is done with 20\% of the Fremont data set and 20\% of the Irvine data set.
The simulated powertrain dynamics takes as inputs the engine torque and switch, the auxiliary power usage, and the demanded torque, and outputs the fuel rate, SOC and engine status.

The powertrain model is then validated using real-world driving data and is shown in Table \ref{table : ModelVal}. 
Table \ref{table : ModelVal} shows the average MPGe over all trips for each route.
The validation result demonstrates that our simulated powertrain dynamics is close to real-driving data.

\begin{table}[hb]
\begin{center}
\caption{Model validation results for Fremont, Irvine, and Highland routes}\label{table : ModelVal}
\begin{tabular}{ccccc}
Routes
& Tot. Miles
& \begin{tabular}[]{@{}c@{}} Measured\\ MPGe\end{tabular} 
& \begin{tabular}[]{@{}c@{}}Simulated\\ MPGe\end{tabular} 
& \begin{tabular}[]{@{}c@{}}Avg. Error\\ (Std.) {[}\%{]}\end{tabular} 
\\ \hline
Fremont  
& 1041
& 61.3                                                    
& 60.9                                                     
& 0.6 (0.6)                                                            \\ \hline
Irvine
& 1270
& 82.4                                                   
& 82.1                                                   
& 0.4 (0.6)                                                          
\\ \hline
Highland 
& 1137
& 82.2                                                 
& 83.0                                                     
& -1.0 (0.6)                                                          \\ \hline
\end{tabular}
\end{center}
\end{table}

\section{Simulation Results and Analysis}
\label{section : result}
The closed loop simulation for the proposed EMS is obtained in four steps: (i) all driving profiles are pre-computed using DP approach and their optimal value functions and state features are stored; (ii) once a target driving profile is selected, the powertrain control training system trains control parameters based on all driving profiles from the same route \textit{except} the target driving profile; (iii) the on-board powertrain control system solves the MPC powertrain control problem, using the approximated value function which is based on the computed control parameters and current state features; (iv) the computed control inputs are sent to the powertrain dynamics, closing the loop. 

To emphasize the effectiveness of the proposed control strategy, an additional predictive EM strategy is simulated, and it is designed as follows: (i) the representative SOC profile is obtained for each route by averaging the desired SOC profiles computed from the DP solutions; (ii) Adaptive ECMS (A-ECMS) strategy is then used to follow the representative SOC profile as in \cite{Yu2011Trip}. The sampling time for the proposed EMS and the A-ECMS is set to 200 ms, and $N_h$ is set to 1 to balance computation with confidence in $\hat{V}$.

\begin{figure}[!t]
    \centering
    \includegraphics[width=0.8\linewidth, height=0.51\linewidth]{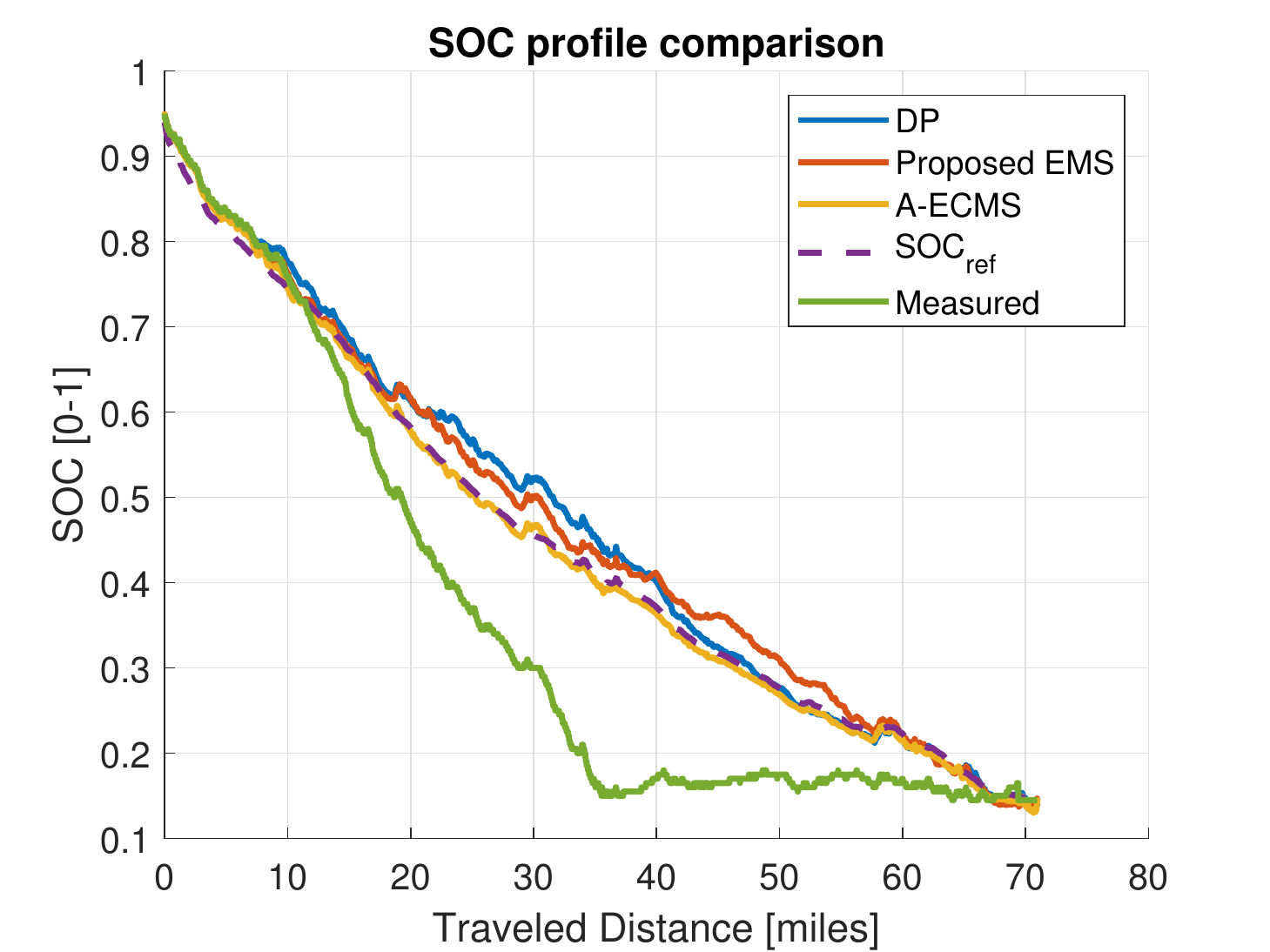}
    \caption{Highland route SOC profile comparison.}
    \label{fig : highland_SOC_profile_example}
\end{figure}

The simulation results for the proposed EMS control design for all collected driving profiles are shown in Fig. \ref{fig : Sim_Results} and Table \ref{table : Final Result}. 
As shown in Fig. \ref{fig : Sim_Results} and Table \ref{table : Final Result}, the proposed EMS improves energy performance by 3.3\%, 7.3\%, and 6.5\% on average for the Fremont, Irvine, and Highland routes respectively, when compared to the baseline EMS (CD-CS). 
Additionally, it improves energy performance by 1.3\%, 1.2\%, and 1.9\%, respectively, when compared to the designed A-ECMS, but it deteriorates performance by 1.3\%, 2.6\%, and 2.0\%, respectively, when compared to the optimal, non-causal EMS computed by DP.
Figure \ref{fig : highland_SOC_profile_example} demonstrates one example of SOC profile comparison for the Highland route (\#5 case in Fig. \ref{fig : Sim_Results}(c)). As we can observe from travel distance around 20 miles to 40 miles, A-ECMS tries to follow its SOC reference trajectory while the proposed EMS exhibits a near-optimal SOC trajectory, which allows the proposed EMS to outperform the designed A-ECMS. However, from travel distance around 42 miles to 58 miles, the proposed EMS deviates away from an optimal trajectory causing deterioration in the energy performance from DP.

In Fig. \ref{fig : Sim_Results}(b), driving profiles from \#1 to \#7 represent afternoon traffic data and from \#8 to \#16 represent morning traffic data. 
The average energy saving in the afternoon scenarios (8.8\% for the proposed EMS, 7.4\% for A-ECMS, and 12.0\% for DP) is slightly higher than in the morning scenarios (5.7\%, 4.4\%, and 7.3\%, respectively). 
A possible explanation for this difference is that, in the morning traffic, the second half of the driving cycle is at high-speed, which is relatively efficient even in charge sustaining mode, thereby reducing the margin for improvement for the proposed EMS.

\begin{table}[hb]
\begin{center}
\caption{The proposed EMS simulation results.}\label{table : Final Result}
\begin{tabular}{ccccc}
Routes   & \begin{tabular}[]{@{}c@{}}Measured\\ MPGe\end{tabular} & \begin{tabular}[]{@{}c@{}}A-ECMS\\ MPGe\end{tabular} & \begin{tabular}[]{@{}c@{}}Data-driven EMS\\ MPGe\end{tabular} & \begin{tabular}[]{@{}c@{}}DP\\MPGe\end{tabular} \\ \hline
Fremont  & 61.3                                                    & 62.5 & 63.4                                                     & 64.3                                                            \\ \hline
Irvine   & 82.4                                                   & 88.2    & 89.4                                               & 92.2                                                          \\ \hline
Highland & 82.2                                                 & 86.3 & 88.1                                                     & 90.0                                                          \\ \hline
\end{tabular}
\end{center}
\end{table}

\section{Conclusions and future work}
\label{section : conclusion}

In this work, a data-driven supervisory energy management strategy (EMS) for plug-in hybrid electric vehicles (PHEVs) is presented. 
The proposed EMS consists of two layers, a cloud layer and an on-board layer. 
In the cloud layer, a global optimization is solved using real-world trip data to learn EMS policy parameters.
Using V2C connectivity, the cloud layer services the policy parameters for a certain route upon requests from the on-board layer. 
The on-board layer computes a real-time solution to the powertrain EMS problem using an MPC scheme parametrized by the learned policy parameters.

The proposed EMS is then evaluated using a high-fidelity powertrain simulation model, which has been validated on real world driving data. 
For collected trip data on three commuting routes in California, the proposed EMS shows 3.3\%, 7.3\%, and 6.5\% improvement in average MPGe when compared to a baseline EMS, and 1.3\%, 1.2\%, and 1.9\% improvement when compared to the designed A-ECMS, and loses only 1.3\%, 2.6\%, and 2.0\% compared to an optimal, non-causal EMS computed by dynamic programming.
Future work includes hardware-in-the-loop experiments and studying the effectiveness of various supervised learning methods. We will also evaluate the effectiveness of the learned policies on routes that we have not traveled yet.
\begin{ack}
This work was funded by the Advanced Research Projects Agency-Energy (ARPA-E), U.S. Department of Energy, under Award Number DE-AR0000791. The views and opinions of authors expressed herein do not necessarily state or reflect those of the United States Government or any agency thereof.
The authors would also like to thank Hyundai America Technical Center, Inc. for providing us with the vehicle information and the real-world trip information used in this paper.
\end{ack}

\bibliography{ifacconf}             

\begin{thebibliography}{19}
\providecommand{\natexlab}[1]{#1}
\providecommand{\url}[1]{\texttt{#1}}
\providecommand{\urlprefix}{URL }
\expandafter\ifx\csname urlstyle\endcsname\relax
  \providecommand{\doi}[1]{doi:\discretionary{}{}{}#1}\else
  \providecommand{\doi}{doi:\discretionary{}{}{}\begingroup
  \urlstyle{rm}\Url}\fi

\bibitem[{Bertsekas(2017)}]{Bertsekas2017DP1}
Bertsekas, D.P. (2017).
\newblock \emph{{Dynamic Programming and Optimal Control Vol. I, 4th ed}}.
\newblock Athena Scientific.

\bibitem[{Cohda(2019)}]{GPSMK5OBU}
Cohda (2019).
\newblock Cohda wireless mk5 obu product brief sheet.
\newblock
  \urlprefix\url{https://www.cohdawireless.com/wp-content/uploads/2018/08/CW_Product-Brief-sheet-MK5-OBU.pdf}.

\bibitem[{Di~Cairano et~al.(2014)Di~Cairano, Bernardini, Bemporad, and
  Kolmanovsky}]{Cairano2014Stochastic}
Di~Cairano, S., Bernardini, D., Bemporad, A., and Kolmanovsky, I.V. (2014).
\newblock {Stochastic MPC with learning for driver-predictive vehicle control
  and its application to HEV energy management}.
\newblock \emph{IEEE Transactions on Control Systems Technology}, 22(3),
  1018--1031.

\bibitem[{(EPA)(2016)}]{EPA2016New}
(EPA), U.E.P.A. (2016).
\newblock New fuel economy and environment labels for a new generation of
  vehicles.
\newblock \emph{EPA}.

\bibitem[{Ericsson(2001)}]{Ericsson2001Independent}
Ericsson, E. (2001).
\newblock {Independent driving pattern factors and their influence on fuel-use
  and exhaust emission factors}.
\newblock \emph{Transportation Research Part D: Transport and Environment}, 6,
  325--345.

\bibitem[{Guanetti et~al.(2016)Guanetti, Formentin, and
  Savaresi}]{Guanetti2016}
Guanetti, J., Formentin, S., and Savaresi, S. (2016).
\newblock {Energy Management System for an Electric Vehicle with a Rental Range
  Extender: A Least Costly Approach}.
\newblock \emph{IEEE Transactions on Intelligent Transportation Systems},
  17(11).

\bibitem[{Guzzella and Sciarretta(2013)}]{Guzzella2013Vehicle}
Guzzella, L. and Sciarretta, A. (2013).
\newblock \emph{{Vehicle Propulsion Systems}}.
\newblock Springer.

\bibitem[{Johannesson et~al.(2007)Johannesson, Asbogard, and
  Egardt}]{Johannesson2007Assessing}
Johannesson, L., Asbogard, M., and Egardt, B. (2007).
\newblock {Assessing the Potential of Predictive Control for Hybrid Vehicle
  Powertrains Using Stochastic Dynamic Programming}.
\newblock \emph{IEEE Transactions on Intelligent Transportation Systems}, 8,
  71--83.

\bibitem[{Manzie et~al.(2015)Manzie, Dewangan, Corde, Grondin, and
  Sciarretta}]{Manzie2015State}
Manzie, C., Dewangan, P., Corde, G., Grondin, O., and Sciarretta, A. (2015).
\newblock {State of Charge Management for Plug-In Hybrid Vehicles With
  Uncertain Trip Information}.
\newblock \emph{Journal of Dynamic Systems, Measurement, and Control}, 137,
  091005.

\bibitem[{Moura et~al.(2011)Moura, Fathy, Callaway, and
  Stein}]{Moura2011Stochastic}
Moura, S.J., Fathy, H.K., Callaway, D.S., and Stein, J.L. (2011).
\newblock {A Stochastic Optimal Control Approach for Power Management in
  Plug-In Hybrid Electric Vehicles}.
\newblock \emph{IEEE Transactions on Control Systems Technology}, 19(3),
  545--555.

\bibitem[{Murgovski et~al.(2012)Murgovski, Johannesson, and
  Sjoberg}]{Murgovski2012Convex}
Murgovski, N., Johannesson, L., and Sjoberg, J. (2012).
\newblock {Convex modeling of energy buffers in power control applications}.
\newblock \emph{IFAC Proceedings Volumes}, 45, 92--99.

\bibitem[{Musardo et~al.(2005)Musardo, Rizzoni, Guezennec, and
  Staccia}]{Musardo2005AECMS}
Musardo, C., Rizzoni, G., Guezennec, Y., and Staccia, B. (2005).
\newblock {A-ECMS: An Adaptive Algorithm for Hybrid Electric Vehicle Energy
  Management}.
\newblock \emph{European Journal of Control}, 11, 509--524.

\bibitem[{Opila et~al.(2012)Opila, Wang, McGee, Gillespie, Cook, and
  Grizzle}]{Opila2012Energy}
Opila, D.F., Wang, X., McGee, R., Gillespie, R.B., Cook, J.A., and Grizzle,
  J.W. (2012).
\newblock {An energy management controller to optimally trade off fuel economy
  and drivability for hybrid vehicles}.
\newblock \emph{IEEE Transactions on Control Systems Technology}, 20(6),
  1490--1505.

\bibitem[{P.~Sharer and Pagerit(2008)}]{Sharer2008Plug}
P.~Sharer, A.~Rousseau, D.K. and Pagerit, S. (2008).
\newblock {Plug-in Hybrid Electric Vehicle Control Strategy: Comparison between
  EV and Charge-Depleting Options}.
\newblock In \emph{SAE Proceedings}.

\bibitem[{Qi et~al.(2016)Qi, Wu, Boriboonsomsin, Barth, and
  Gonder}]{Qi2016Data}
Qi, X., Wu, G., Boriboonsomsin, K., Barth, M., and Gonder, J. (2016).
\newblock {Data-Driven Reinforcement Learning–Based Real-Time Energy
  Management System for Plug-In Hybrid Electric Vehicles}.
\newblock \emph{Journal of the Transportation Research Board}, 2572, 1--8.

\bibitem[{Serrao et~al.(2011)Serrao, Onori, and
  Rizzoni}]{Serrao2011Comparative}
Serrao, L., Onori, S., and Rizzoni, G. (2011).
\newblock {A Comparative Analysis of Energy Management Strategies for Hybrid
  Electric Vehicles}.
\newblock \emph{Journal of Dynamic Systems, Measurement, and Control}, 133,
  031012.

\bibitem[{Stockar et~al.(2011)Stockar, Marano, Canova, Rizzoni, and
  Guzzella}]{Stockar2011Energy}
Stockar, S., Marano, V., Canova, M., Rizzoni, G., and Guzzella, L. (2011).
\newblock {Energy-optimal control of plug-in hybrid electric vehicles for
  real-world driving cycles}.
\newblock \emph{IEEE Transactions on Vehicular Technology}, 60, 2949--2962.

\bibitem[{Stockar et~al.(2009)Stockar, Tulpule, Marano, and
  Rizzoni}]{Stockar2009Energy}
Stockar, S., Tulpule, P., Marano, V., and Rizzoni, G. (2009).
\newblock {Energy, Economical and Environmental Analysis of Plug-In Hybrids
  Electric Vehicles Based on Common Driving Cycles}.
\newblock \emph{SAE International Journal of Engines}, 2(2), 2009--24--0062.

\bibitem[{{Yu} et~al.(2011){Yu}, {Kuang}, and {McGee}}]{Yu2011Trip}
{Yu}, H., {Kuang}, M., and {McGee}, R. (2011).
\newblock Trip-oriented energy management control strategy for plug-in hybrid
  electric vehicles.
\newblock In \emph{2011 50th IEEE Conference on Decision and Control and
  European Control Conference}, 5805--5812.

\end{thebibliography}

\end{document}